\documentclass[fp,twocolumn]{jpsj3}
\usepackage{txfonts}
\usepackage{bm}
\usepackage{cite}
\usepackage{amsmath}
\newcommand{\tn}{$T_{\rm m}$}
\usepackage{comment}
\usepackage{mediabb}
\usepackage{color}
\usepackage{graphicx}

\title{Uncompensated Antiferromagnetic Ordering of UAu$_2$Si$_2$ Studied by $^{29}$Si-NMR}

\author{Chihiro Tabata\thanks{ctabata@post.kek.jp}\thanks{Present address: Condensed Matter Research Center and Photon Factory,
Institute of Materials Structure Science, High Energy Accelerator Research Organization, Tsukuba, Japan}, Yoshihiko Ihara, Shogo Shimmura,  Naoyuki Miura, Hiroyuki Hidaka, Tatsuya\nobreak Yanagisawa, and Hiroshi Amitsuka}
\inst{Graduate School of Science, Hokkaido University, Sapporo 060-0810, Japan \\
} 

\abst{The magnetically ordered phase of body-centered tetragonal UAu$_2$Si$_2$ was explored through $^{29}$Si-NMR experiments. The field-swept NMR spectra show asymmetric peak splitting below magnetic phase transition temperature, $T_{\rm m}$ = 20 K. This splitting is well explained by the occurrence of a magnetic order with propagation vector of $\bm{q}$ = (2/3, 0, 0) and magnetic moments pointing parallel to the tetragonal [001] axis, offering evidence for uncompensated ``up-up-down''-type antiferromagnetic ordering with a spontaneous [001] magnetization component. A symmetry analysis of the hyperfine-coupling tensor with phenomenological corrections including orbital and dipolar-field contributions yields the magnitude of the ordered moment to be approximately 1.4 $\pm$ 0.2 Bohr magneton per uranium ion, and evaluates the site-dependent hyperfine-coupling constants. } 

\begin{document}
\maketitle

\section{Introduction}
For recent three decades, uranium intermetallic compounds have been studied intensively, particularly from the aspects of magnetism and superconductivity. Among them, a series of U$T_2$Si$_2$ ($T$: transition metal elements) compounds has been one of the most investigated groups of substances. They show a variety of low-temperature properties, such as commensurate-incommensurate magnetic phase transitions, heavy-electron states, superconductivity, hidden order, etc. 
Systematic studies of this series of compounds have still actively been performed to clarify the controlling role of the $T$-elements in those varieties of low-temperature properties and to gain a better understanding of 5f-electronic states in metallic uranium compounds.

In the present study, we focus on UAu$_2$Si$_2$, which crystallizes in ThCr$_2$Si$_2$ type body-centered tetragonal (bct) structure (space group: $I4/mmm$, no. 139, $D^{17}_{4h}$). Previous studies made on polycrystalline samples all suggested that this compound exhibits a ferromagnetic order below around 19 K \cite{Palstra85, Saran88, Rebelsky91, Torikachvili92, Lin97}. 
On the other hand, we have recently suggested from specific heat and magnetization measurements on single-crystalline samples that the order is much more likely to be of uncompensated antiferromagnetic (AFM) than simply of ferromagnetic (FM) \cite{Tabata16}.
The transition temperature is estimated to be $\sim$ 20 K, which we refer to as $T_{\rm m}$ hereafter.
We also observed a parasitic ferromagnetic component, which develops below $\sim$50 K.
Another remarkable feature of UAu$_2$Si$_2$ would be an enhancement of the electronic specific-heat coefficient $\gamma$, which is evaluated to be $\sim$ 180 mJ/K$^2$mol even in the ordered state. This implies that the heavy electrons coexist with the ordered magnetic moments in this system. The coexistence between the heavy-electron state and a well-defined second-order phase transition is similarly observed in URu$_2$Si$_2$, whose ordered phase has been a subject of controversy for several decades as ``hidden order''. Although the nature of the order is considered to be different, it might be expected to get insights to describe the duality of 5f electrons in URu$_2$Si$_2$ from the detailed study on UAu$_2$Si$_2$. In the present paper, we report $^{29}$Si-NMR experiments performed on polycrystalline UAu$_2$Si$_2$ in order to obtain microscopic information about its magnetic properties, particularly the static magnetic structure.  

\section{Experiment}
The polycrystalline sample was prepared by the arc-melting method in Ar atmosphere and annealed in vacuum at 900$^{\circ}$C for 7 days. Powder X-ray diffraction measurements confirmed that the obtained crystal has the ThCr$_2$Si$_2$-type bct structure with lattice parameters of $a = 4.222(1)$ \AA \ and  $c = 10.29(1)$ \AA. 
The impurity phases were not detected within the accuracy of 0.5 \% of the main-peak intensity.
For the NMR measurements, the sample was crushed into powder and mixed with stycast 1266. Then it was solidified in a magnetic field of $\sim$ 1 T so that the powdered sample became fixed in a magnetically oriented alignment. The field-sweep NMR spectra were taken by the spin-echo method in a temperature range from 70 K to 4.2 K. We chose a frequency of 38.5 MHz through the present measurements. The external field was calibrated from $^{63}$Cu resonance. The field-sweep range was selected in the region of a low-field ordered phase I on the $H$-$T$ phase diagram in Ref. \citen{Tabata16}, which ensures no field-induced phase transition during sweeping the external field at 4.2 K. 

\section{Results and Analysis}
\begin{figure}[ht]
	\begin{center}
		\includegraphics[width=8cm]{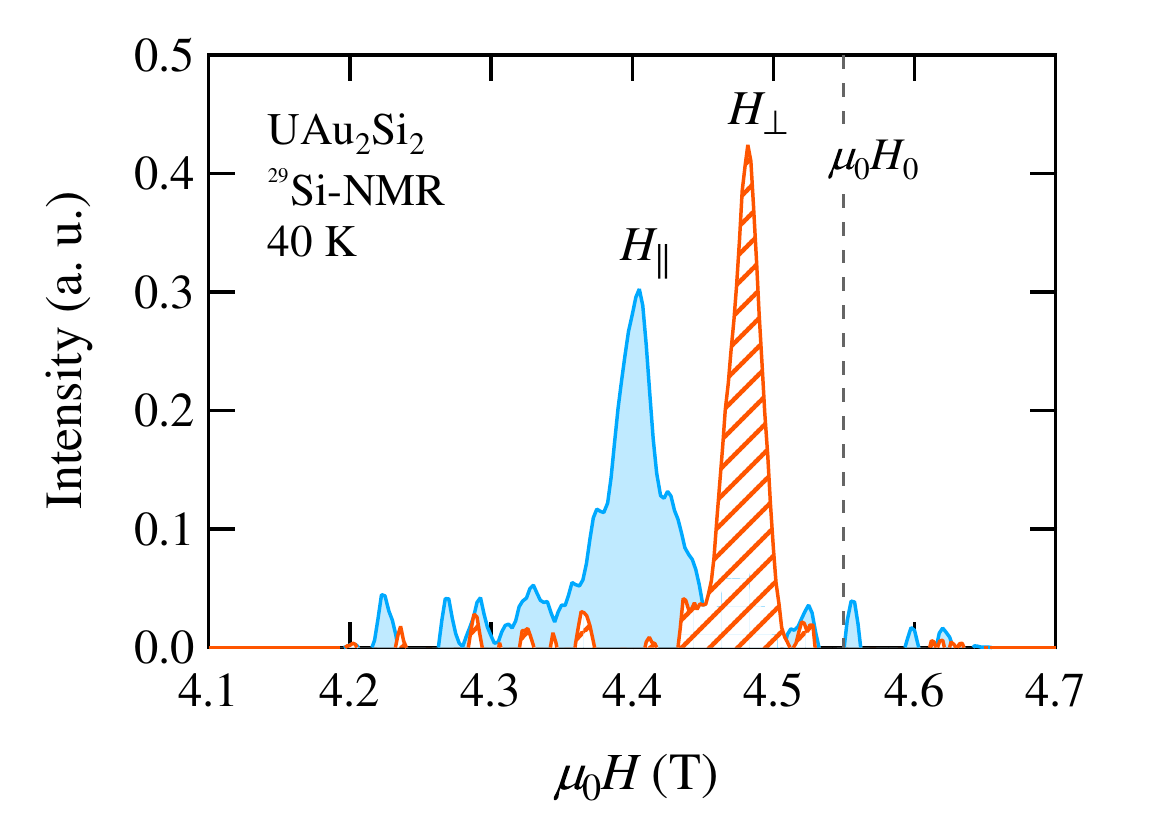}
	\end{center}
	\caption{(Color online) Typical spectra of $^{29}$Si-NMR of the magnetically oriented powdered sample of UAu$_2$Si$_2$ measured at 40 K in the magnetic fields parallel and perpendicular to the [001] axis.}
	\label{NMR_spectra_ac}
\end{figure}

Figure \ref{NMR_spectra_ac} exhibits typical $^{29}$Si-NMR spectra obtained by sweeping the magnetic field parallel ($H_{\parallel}$) and perpendicular ($H_{\perp}$) to the magnetically-easy axis, namely, the [001] axis. 
Since $^{29}$Si nuclei have spin 1/2, the NMR spectra have a single peak in a paramagnetic state. 
In this condition, the resonance field of the free Si nuclear spin is calculated as ${\mu}_{0}H_{\rm 0} = 38.5/\gamma \sim 4.55$ T, where $\gamma$ is the gyromagnetic ratio of $^{29}$Si.
The observed peaks of NMR spectra in the fields of $H_\perp$ and  $H_\parallel$ are well separated to each other, indicating that the present powdered sample is successfully oriented. 

\begin{figure}[ht]
	\begin{center}
		\includegraphics[width=8cm]{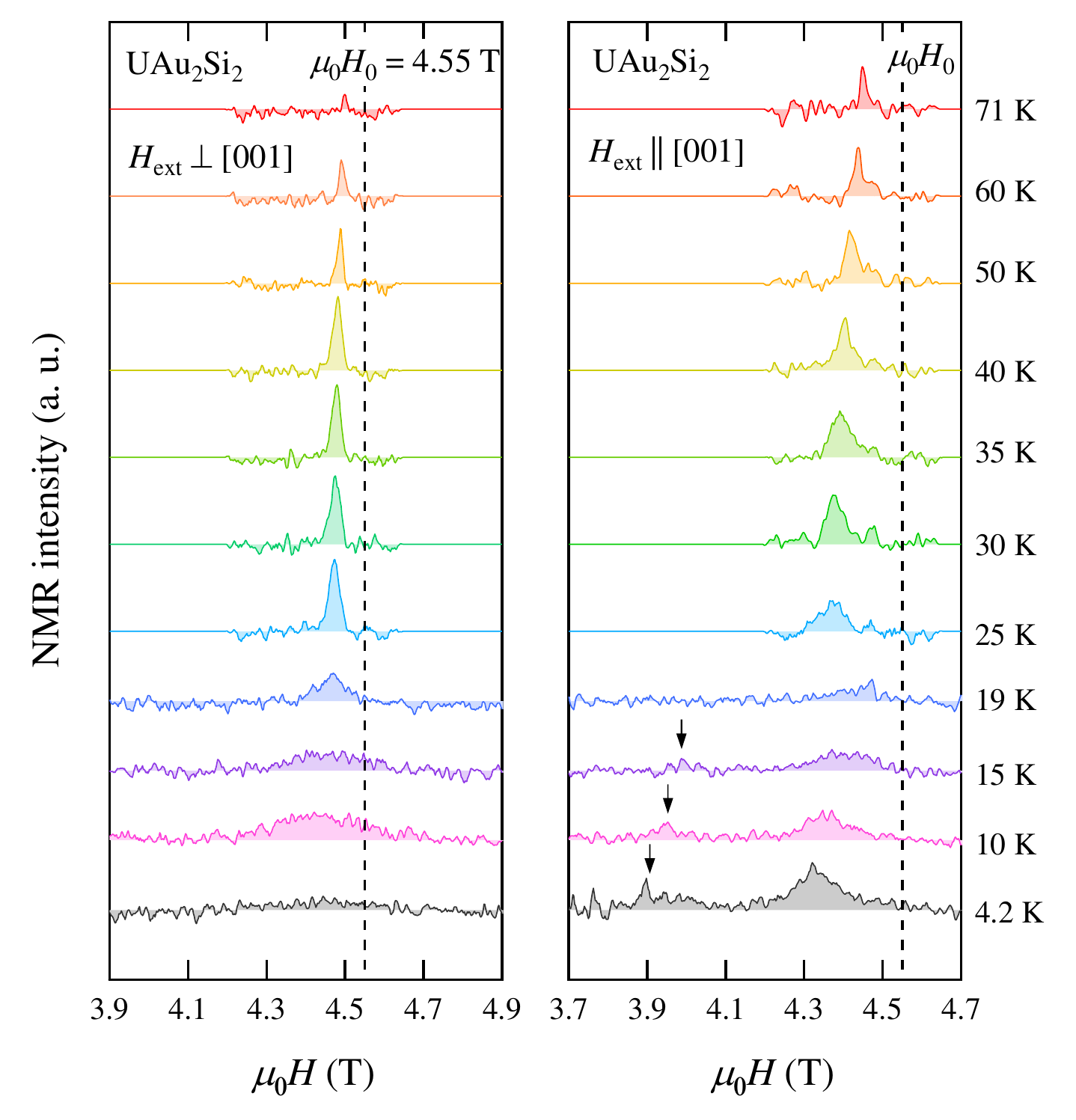}
	\end{center}
	\caption{(Color online) Temperature dependences of $^{29}$Si-NMR spectra of the UAu$_2$Si$_2$ powdered polycrystalline sample, measured in external magnetic fields perpendicular (left) and parallel (right) to [001]. The broken line denotes the resonance field expected for a free Si atom: $\mu_{\rm 0}H_{\rm 0} = 4.55$ T.}
	\label{NMR_spectra_all}
\end{figure}

\begin{figure}[b]
	\begin{center}
		\includegraphics[width=8cm]{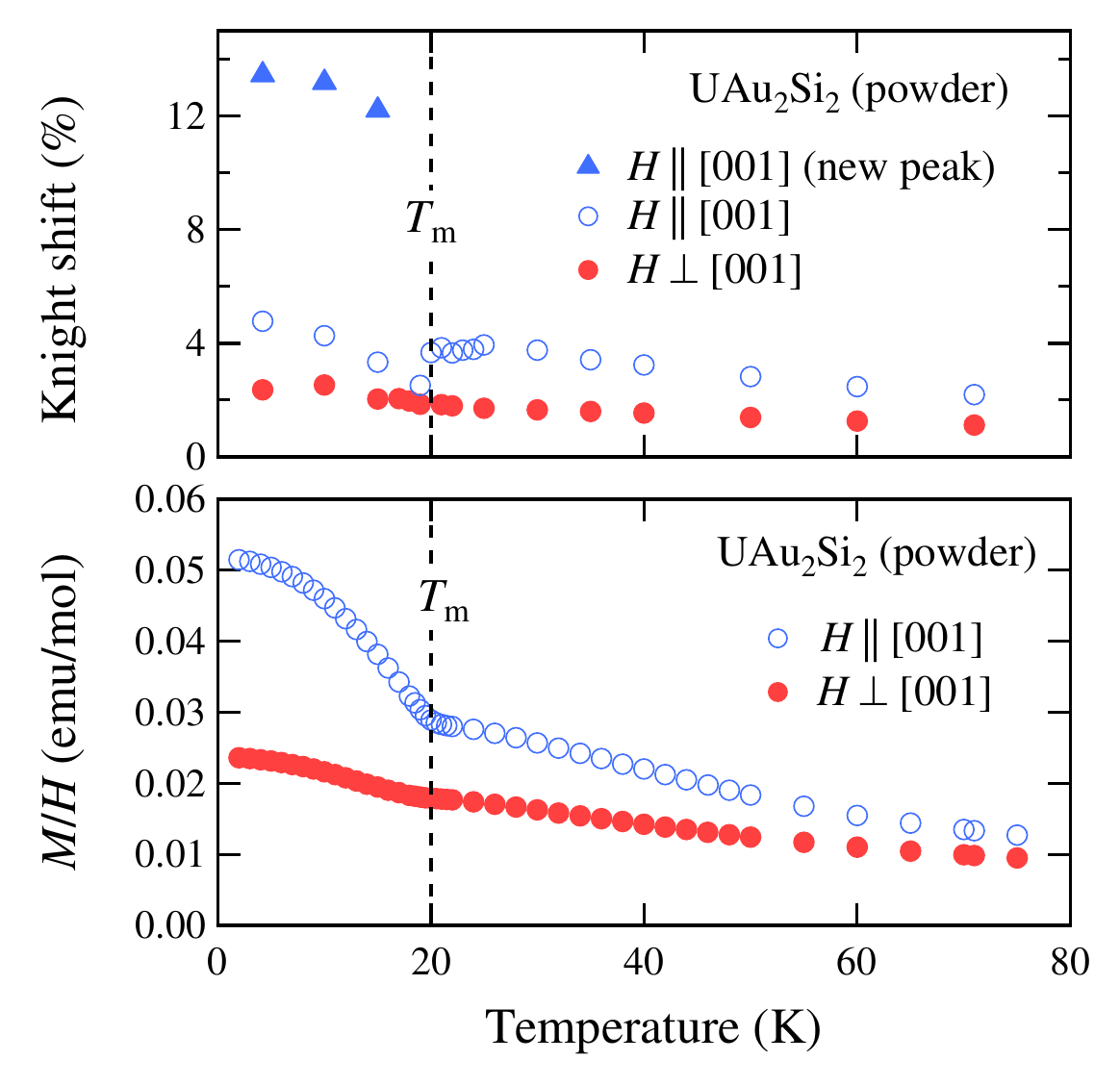}
	\end{center}
	\caption{(Color online) Temperature dependences of the Knight shift (top) and magnetic susceptibility (bottom) of the oriented powdered sample. The data labeled as ``new peak'' indicate the Knight shift evaluated from the NMR peak marked by arrows in Fig. \ref{NMR_spectra_all}.}
	\label{K_and_chi_vs_T}
\end{figure}

\begin{figure}[h]
	\begin{center}
		\includegraphics[width=8cm]{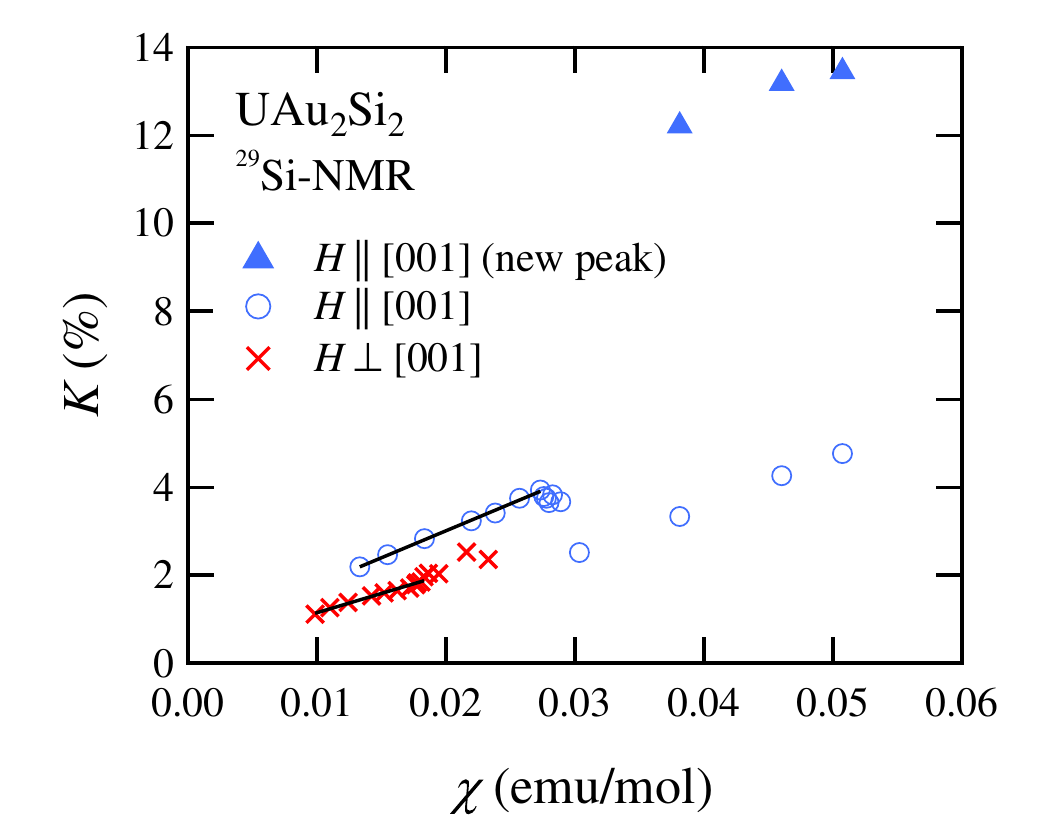}
	\end{center}
	\caption{(Color online) Knight shift vs. magnetic susceptibility of the oriented powdered UAu$_2$Si$_2$. The solid line is the least-square fit within the paramagnetic state.}
	\label{K_chi}
\end{figure}

Figure \ref{NMR_spectra_all} shows the temperature dependence of the $^{29}$Si-NMR spectra measured at temperatures ranging from 71 K down to 4.2 K. 
In the paramagnetic state, the resonance peaks are observed in magnetic fields lower than $\mu_{0}H_{0}$, meaning a positive Knight shift.
Below \tn, we observed a distinct difference in the spectra between the two field directions: for $H_{\parallel}$ a small new peak appears in the lower field region as indicated by the arrows in Fig. \ref{NMR_spectra_all},  while for $H_\perp$ the resonance peak just becomes broader. 
This is direct evidence that an AFM order occurs with at least two magnetic sublattices and is consistent with a cusp anomaly of magnetization observed on a single crystal in the low field range \cite{Tabata16}.
The transverse and longitudinal components of Knight shift, $K_\perp$ and $K_\parallel$, were evaluated at the center of resonance peaks determined by fittings using the Gaussian function. 
Plotted in Fig. \ref{K_and_chi_vs_T} are temperature dependences of $K_\perp$ and $K_\parallel$.
They show similar anisotropy to that observed in the magnetization of a powdered sample prepared in the same manner as described above. 
We observed linear relations between $K_\perp$ and $K_\parallel$ and the magnetic susceptibilities above $T_{\rm m}$, as shown in Fig. \ref{K_chi}. 
From the linear fitting of these $K$-$\chi$ plots, the spin hyperfine coupling constants are estimated as $A_{\parallel} \sim$ 0.69 T/${\mu}_{\rm B}$ and $A_{\perp} \sim$ 0.48 T/${\mu}_{\rm B}$. They are of the same order of magnitude as those reported in the isostructural relatives: $A_{\rm isotropic} \sim$ 0.36 T/${\mu}_{\rm B}$ of URu$_2$Si$_2$ \cite{Kohori96}, $A_{ab} \sim$ 0.28 T/${\mu}_{\rm B}$ of CePd$_2$Si$_2$ \cite{Kawasaki98}. 

\begin{figure}[h]
	\begin{center}
		\includegraphics[width=8.0cm]{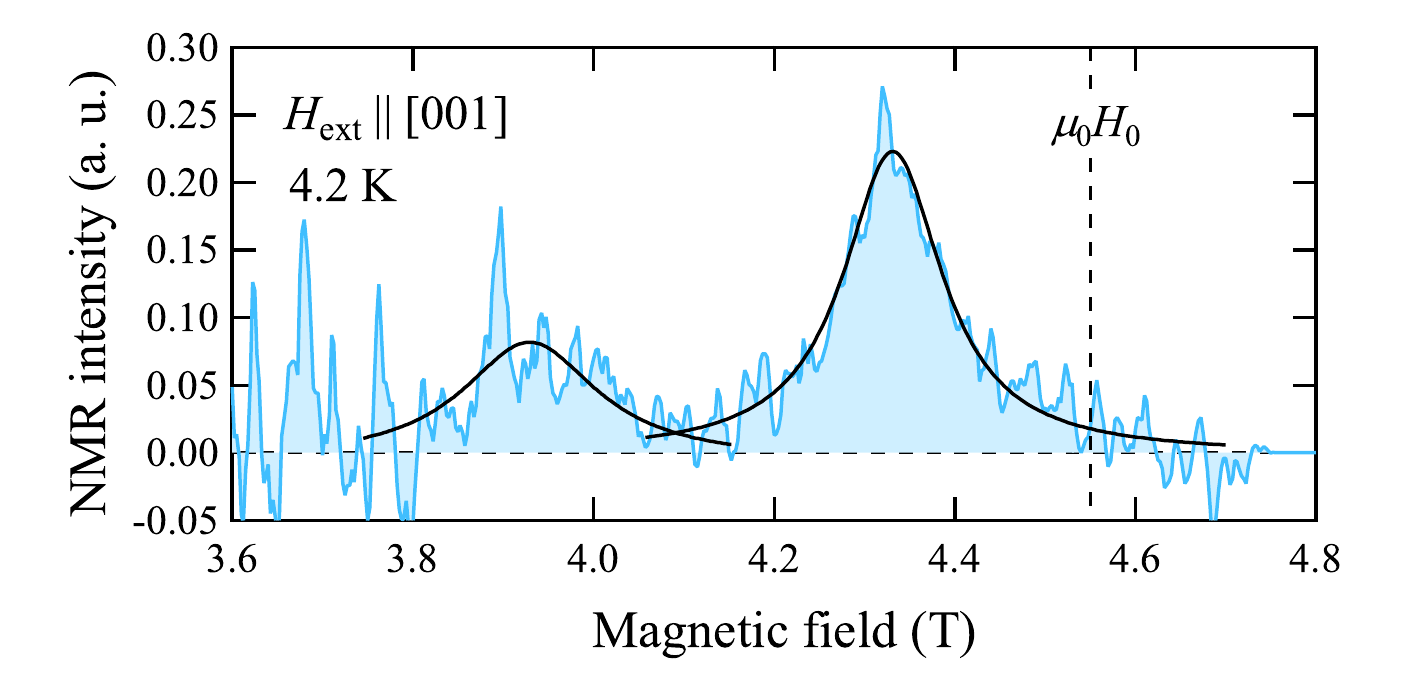}
	\end{center}
	\caption{(Color online) $^{29}$Si-NMR spectra of the UAu$_2$Si$_2$ powdered sample measured at 4.2 K for fields parallel to the [001] axis. Solid curves indicate the best fits of the data using the Lorentzian function with parameters summarized in Table \ref{tab:NMR_fit_result}.}
	\label{peak_fit_NMR}
\end{figure}

The peak profile measured at 4.2 K in $H_\parallel$ was analyzed by using the lorentzian fitting to evaluate the intensity of each peak as shown in Fig. \ref{peak_fit_NMR}. 
This analysis yields the intensity ratio of $I_{\rm small} : I_{\rm large} \sim 1 : 2$, where $I_{\rm small}$ and $I_{\rm large}$ are integrated intensities of the peaks at $\sim$ 3.9 T and $\sim$ 4.3 T, respectively. 
The parameters obtained by the fitting are listed in Table \ref{tab:NMR_fit_result}. 
The intensity ratio of 1 : 2 means that there are two kinds of Si sites with inequivalent components of internal fields projected on the [001] axis in the ordered state.
This strongly suggests that an AFM order occurs with an ordering period which is three times as long as a period of the primitive crystal lattice. 
In addition, the large broadening of the spectra in $H_\perp$ below $T_{\rm m}$ indicates that there are finite internal fields generated in the basal plane by a magnetic phase transition. 
It is clear that this peak broadening cannot be explained only by the effects of misorientation because the mixing of the $c$-component of the internal field cannot give rise to the intensity for $H > H_{\rm 0}$, contrary to the present observation. 

\renewcommand{\arraystretch}{1.5} 
\begin{table}[h]
\begin{center}
\caption{Magnetic field of the peak center, $H_{\rm reso}$, the mean value of internal fields, ${\Delta}H (\equiv H_{\rm 0} - H_{\rm reso})$, and the full width at half maximum (FWHM), the signal intensity, $I_{\rm reso}$, obtained by the fitting analysis of the $^{29}$Si-NMR spectra taken at 4.2 K.\\}
\begin{tabular}{lllll}
\hline
peak notation  & $H_{\rm reso} (T)$  & ${\Delta}H$ (T) & FWHM (T) & $I_{\rm reso}$ (a. u. )\\
\hline
small peak       & 3.94            & 0.62       & 0.18    & 0.016           \\
large peak       & 4.33            & 0.22       & 0.13    & 0.030             \\
\hline 
\end{tabular}
\label{tab:NMR_fit_result}
\end{center}
\end{table}

\begin{figure}
\centering
		\includegraphics[width=7cm]{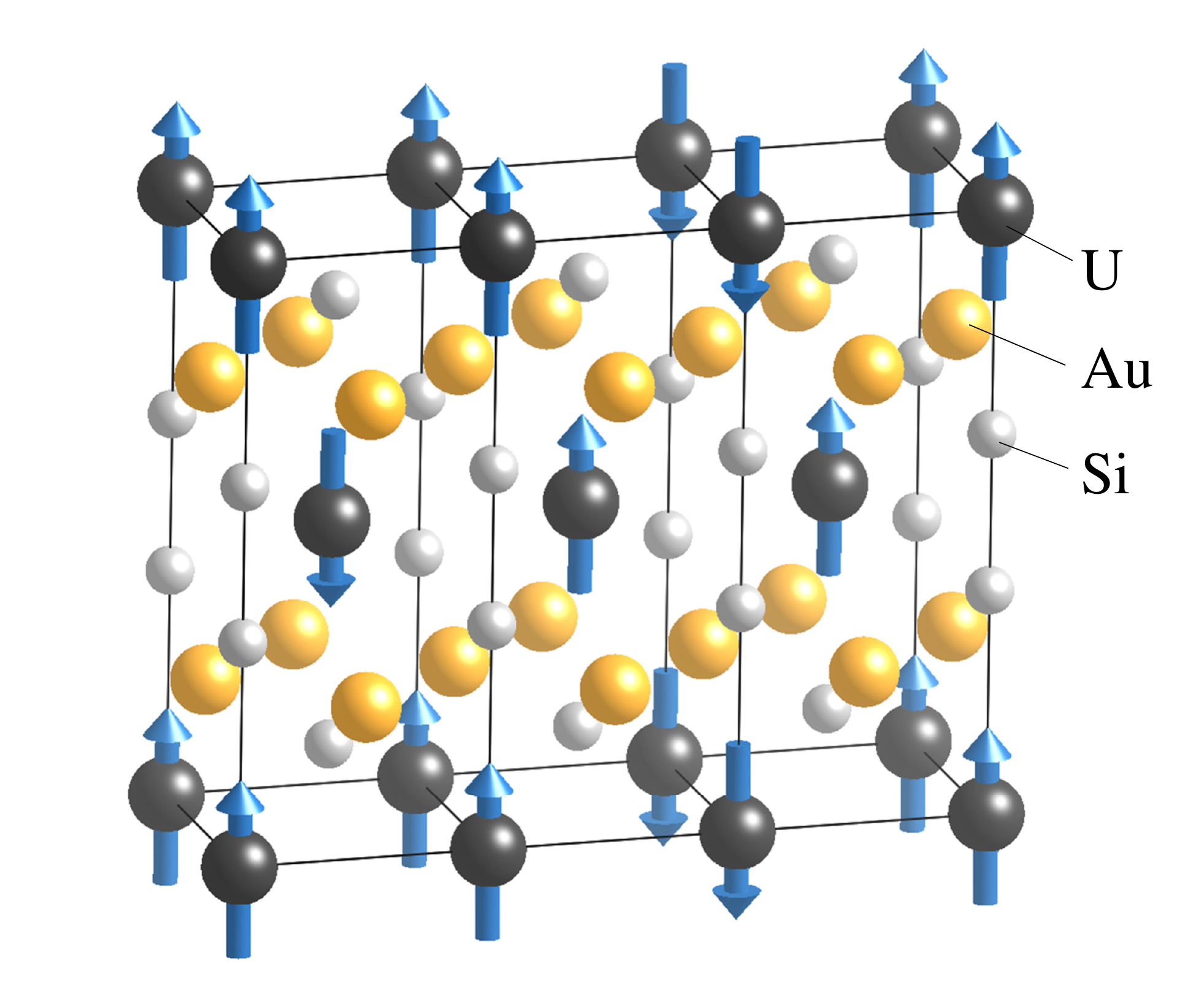}
	\caption{(Color online) Magnetic structure proposed from the $^{29}$Si-NMR experiments. The propagation vector $\bm{q}$ is (2/3, 0, 0), and the magnetic moments are along the [001] axis.}
	\label{UAu2Si2_spins}
\end{figure}

\begin{figure}[h]
\centering
		\includegraphics[width=7.5cm]{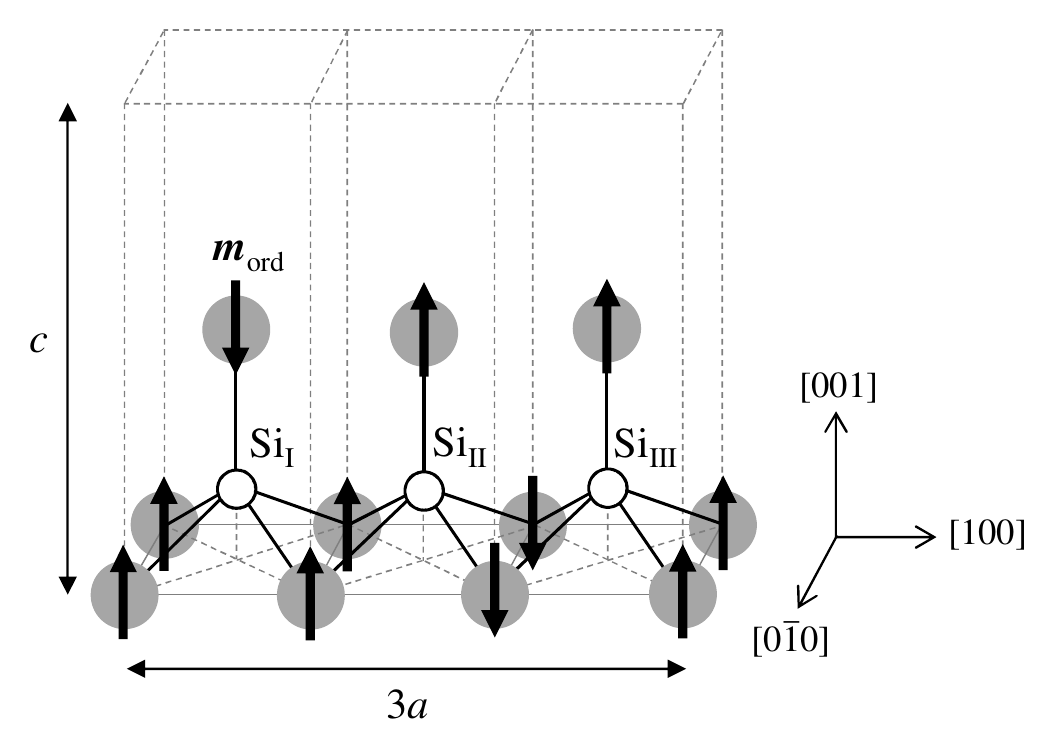}
	\caption{Schematic view of three magnetically-inequivalent Si sites produced by the uncompensated up-up-down AFM state with $\bm{q} = (2/3, 0, 0)$.}
	\label{sites_Si}
\end{figure}

Now let us discuss the magnetic structure to account for the NMR spectra observed in the ordered state of UAu$_2$Si$_2$.
The present data show that there are two kinds of Si magnetic sites at which the [001] components of the internal fields are inequivalent, and the ratio of the site number is about 1 : 2. 
Combined these with the fact that the spontaneous magnetization has only the [001] component in our magnetization measurements \cite{Tabata16}, we have concluded that the simplest candidate of the magnetic structure can be an up-up-down configuration of magnetic moments pointing along [001].
Then, the remaining question is the direction of the $\bm{q}$-vector.
A simple consideration of the number of magnetically equivalent Si sites will lead us to the consequence that the uncompensated AFM order with $\bm{q} = (2/3, 0, 0)$ is the most likely magnetic structure of this system.
The magnetic structure is illustrated in Fig. \ref{UAu2Si2_spins}.
In this case, there are three magnetically inequivalent Si sites, which we define as Si$_{\rm I}$, Si$_{\rm II}$, and Si$_{\rm III}$ as shown in Fig. \ref{sites_Si}. 
The internal fields at two of them (Si$_{\rm II}$ and Si$_{\rm III}$) have an identical [001] component.
Hence the NMR peaks for these two sites are shifted by the same magnitude in $H_{\parallel}$, and observed as a single peak with the intensity of twice as large as that for Si$_{\rm I}$.
Since the hyperfine fields generated by the four spins sitting at the first-nearest uranium ions are canceled out at Si$_{\rm II}$ and Si$_{\rm III}$, the large peak is expected to be observed nearer to $H_{\rm 0}$ than the small one that originates from Si$_{\rm I}$. 
These features are obviously consistent with the observed NMR spectra below $T_{\rm m}$. 
In contrast, if the $\bm{q}$-vector is parallel to any of [001], [101], and [111], the NMR spectrum splits into three peaks in the ordered state.
On the other hand, if $\bm{q} \parallel [110]$, the number of peaks becomes two. However, the peak intensity ratio is reversed, contrary to the observation; i.e., the peak located farther from $H_{\rm 0}$ has the larger intensity.

\begin{figure}[ht]
\centering
		\includegraphics[width=7cm]{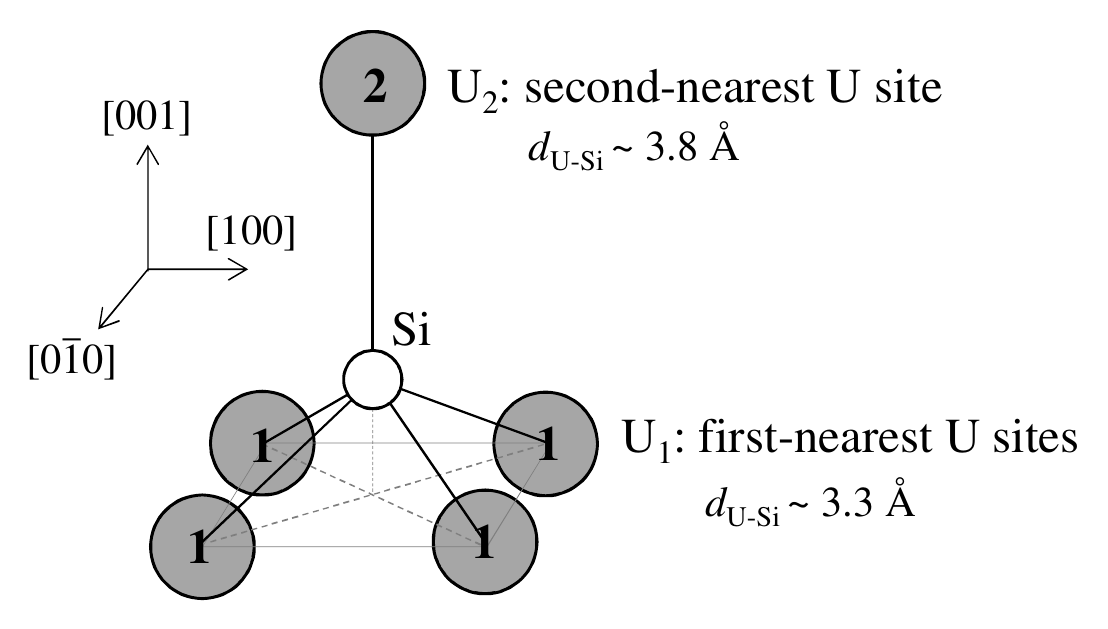}
	\caption{Schematic view of a Si site and its neighboring U sites in UAu$_2$Si$_2$.}
	\label{sites_U}
\end{figure}

We now present a symmetry analysis to estimate the magnitude of the ordered moment and the hyperfine coupling constants, following the previous studies made, for example, on the AFM states of BaFe$_2$As$_2$\cite{Kitagawa08}, LaFeAsO$_{1-x}$F$_{x}$\cite{Kitagawa10}, and LaCoGe\cite{Karube13}.
The internal field $\bm{H}_{\rm int}$ which acts on the Si nuclear spin can be described as follows:
\begin{equation}
\bm{H}_{\rm int} = \bm{H}_{\rm hf} + \bm{H}_{\rm dip} + \bm{H}_{\rm dia},
\end{equation}
where $\bm{H}_{\rm hf}$, $\bm{H}_{\rm dip}$, and $\bm{H}_{\rm dia}$ are the hyperfine field, the dipolar field produced by the local magnetic moments, and the diamagnetic field, respectively. 
We consider that the observed NMR spectra are governed mainly by the short-range transferred hyperfine interactions between Si nucleus and the ordered moments on the four first-nearest-neighbor and one second-nearest-neighbor U sites. 
We refer to these U sites as U$_1$ and U$_2$, respectively, as shown in Fig. \ref{sites_U}. Thus $\bm{H}_{\rm hf}$ can be written as
\begin{equation}
\bm{H}_{\rm hf} = \sum_{i}^{4}A_{i}^{\rm U_1}\bm{m}_{i}^{\rm U_1} + A^{\rm U_2}\bm{m}^{\rm U_2},
\label{eq:Hhf}
\end{equation}
where $A_{i}^{\rm U_1}$ ($i = 1, 2, 3, 4$, which is a label identifying the four ions at U$_1$) and $A^{\rm U_2}$ are hyperfine coupling tensors between a Si nuclear moment and ordered 5f-electron moments $\bm{m}_{i}^{\rm U_1}$ and $\bm{m}^{\rm U_2}$ on the U ions at U$_1$ and U$_2$, respectively.
The internal magnetic fields at the three Si sites (Fig. \ref{sites_Si}) in the ordered state with $\bm{q}$ = (2/3, 0, 0) can be described as follows:
\begin{align}
\begin{aligned}
&\bm{H}_{\rm int}^{\rm Si_{I}} 
= m_{\rm ord}\begin{bmatrix}
0\\
0\\
4A_{cc}^{\rm U_1}-A_{cc}^{\rm U_2}
\end{bmatrix} + 
\chi^{\rm orb}\bm{H}_{\rm ext} A^{\rm orb} +
m_{\rm ord}\begin{bmatrix}
0\\
0\\
-0.10
\end{bmatrix}, \\
&\bm{H}_{\rm int}^{\rm Si_{II}} = m_{\rm ord}
\begin{bmatrix}
4A_{ac}^{\rm U_1}\\
0\\
A_{cc}^{\rm U_2}
\end{bmatrix} + 
\chi^{\rm orb} \bm{H}_{\rm ext} A^{\rm orb} +
m_{\rm ord}\begin{bmatrix}
0.091\\
0\\
-0.030
\end{bmatrix}, \\
{\rm and} \\
&\bm{H}_{\rm int}^{\rm Si_{III}} = m_{\rm ord}
\begin{bmatrix}
-4A_{ac}^{\rm U_1}\\
0\\
A_{cc}^{\rm U_2}
\end{bmatrix} + 
\chi^{\rm orb} \bm{H}_{\rm ext} A^{\rm orb} +
m_{\rm ord}\begin{bmatrix}
-0.091\\
0\\
-0.030
\end{bmatrix},
\label{eq:Hsite2}
\end{aligned}
\end{align}
where $m_{\rm ord} = \left|\bm{m}_{i}^{\rm U_1}\right| = \left|\bm{m}^{\rm U_2}\right|$, $A_{cc}^{\rm U_1} =\left| A_{i, cc}^{\rm U_1}\right|$, and $A_{ac}^{\rm U_1} = \left|A_{i, ac}^{\rm U_1}\right|$.

In each of three equations, the first term is the hyperfine field derived from Eq. (\ref{eq:Hhf}). $A_{cc}^{\rm U_1}$, $A_{ac}^{\rm U_1}$, and $A_{cc}^{\rm U_2}$ are components of the hyperfine coupling tensors for U$_1$ and U$_2$. 
The second term is the phenomenological correction term of the hyperfine field due to the temperature-independent contributions, such as orbital part, Pauli paramagnetism, van-Vleck paramagnetism, and so on. 
$\chi^{\rm orb}$ and $A^{\rm orb}$ are the temperature-independent magnetic susceptibility tensor and the hyperfine coupling tensor, respectively, both of which cause the temperature-independent NMR shift. $\bm{H}_{\rm ext}$ is an external magnetic field.
The third term is the dipolar field in the unit of Tesla (T) produced by the local ordered spins in the unit of Bohr magneton ($\mu_{\rm B}$), where we calculated numerically the magnetic field strength from the U ions located within a 20 {\AA} radius from the each Si site.
The contribution of diamagnetic field is small and can be neglected in the present calculations.  

\begin{figure}[ht]
\centering
		\includegraphics[width=7cm]{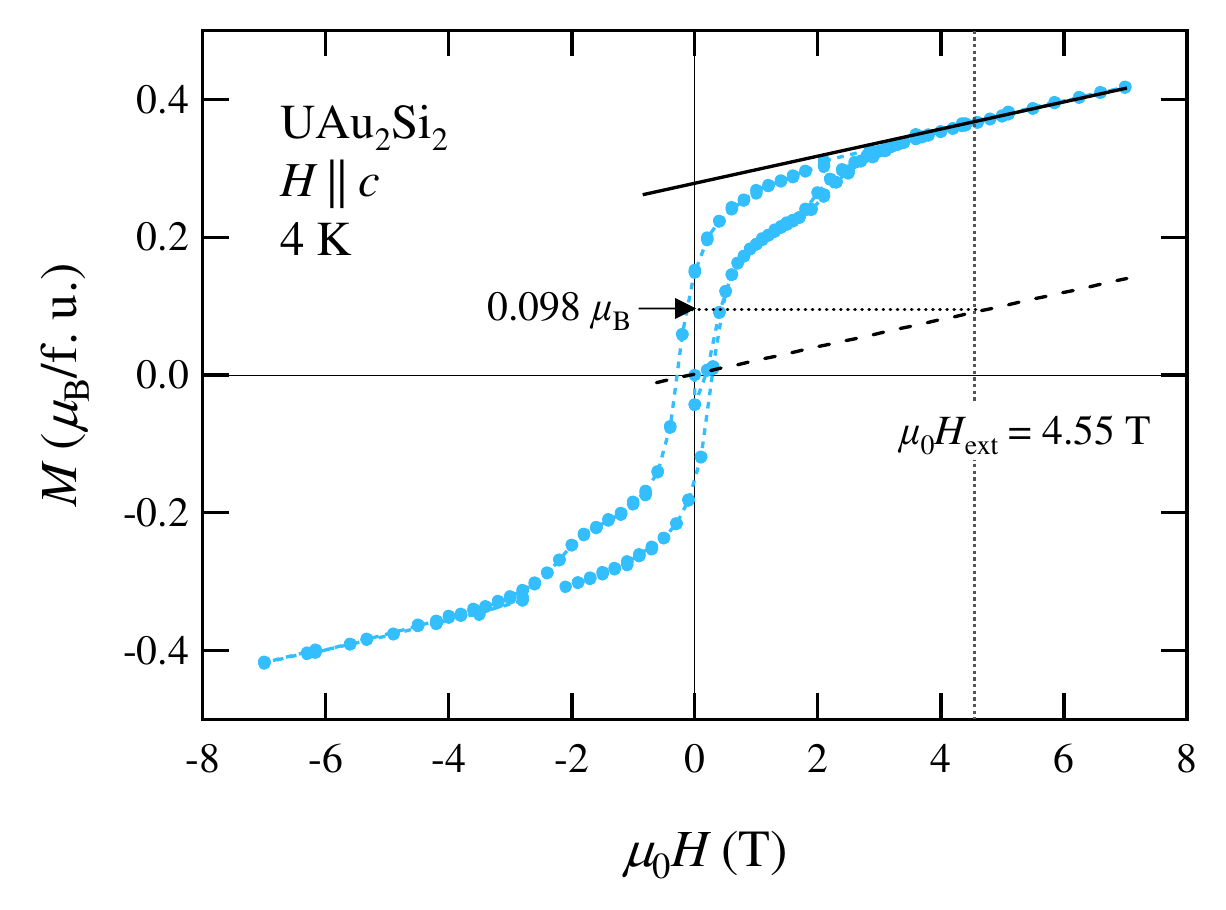}
	\caption{(Color online) Magnetic-field dependence of magnetization along the [001] axis of single-crystalline UAu$_2$Si$_2$, measured at 4 K. The details of the data are given in Ref. \citen{Tabata16}. The solid line denotes the best fitting line for the magnetization curve above 4 T and its extrapolation to lower fields. The broken line is given by translating the solid line to pass it through the origin. The dotted lines are guides to the eye.}
	\label{M_vs_H_4K}
\end{figure}

Figure \ref{M_vs_H_4K} shows a magnetization curve observed at 4 K for $H \parallel [001]$ on a single-crystalline UAu$_2$Si$_2$ \cite{Tabata16}. It is clearly seen that there is a non-negligible $H$-linear component of the $c$-axis magnetization, which can make a significant contribution to the second term in Eqs. (\ref{eq:Hsite2}) through the tensor components $\chi_{cc}^{\rm orb}$ and $A_{cc}^{\rm orb}$ when the external field is applied parallel to [001].
$\chi_{cc}^{\rm orb}$ at 4 K is estimated from the slope of a linear component of the observed $M$-$H$ curve, as $\chi_{cc}^{\rm orb} = 0.098 \ {\mu}_{\rm B} /4.55 \ {\rm T} \sim 0.0215$ $\mu_{\rm B}$/T, as displayed in Fig. \ref{M_vs_H_4K}. On the other hand, $A_{cc}^{\rm orb}$ appears in a relation between the Knigh shift and the susceptibility:
\begin{equation}
K_{\parallel} = A_{cc}^{\rm spin}\chi_{cc}^{\rm spin}(T) + A_{cc}^{\rm orb}\chi_{cc}^{\rm orb},
\label{eq:K-chi-separate}
\end{equation}
where $\chi_{cc}^{\rm spin}(T)$ is the spin part of the magnetic susceptibility, which causes a temperature-dependent NMR shift, and $A_{cc}^{\rm spin}$ is the corresponding hyperfine-coupling constant.
Now we deform the Eq. (\ref{eq:K-chi-separate}) into
\begin{equation}
K_{\parallel} = A_{cc}^{\rm spin}(\chi_{cc}^{\rm spin}(T) + \chi_{cc}^{\rm orb}) + (A_{cc}^{\rm orb}-A_{cc}^{\rm spin})\chi_{cc}^{\rm orb},
\label{eq:K-chi-deform}
\end{equation}
so that it can be compared with the observation. The slope and the intersection obtained using a linear regression analysis for the $K$-$\chi$ plot (Fig. \ref{K_chi}) yield
\begin{equation}
A_{cc}^{\rm spin} = 0.69 \ {\rm T}/{\mu}_{\rm B},
\label{eq:slope}
\end{equation}
and
\begin{equation}
(A_{cc}^{\rm orb}-A_{cc}^{\rm spin})\chi_{cc}^{\rm orb} = 0.55
\label{eq:intersection}
\end{equation}
in Eq. (\ref{eq:K-chi-deform}).
Since we have already obtained the value of $\chi_{cc}^{\rm orb}$, we can estimate $A_{cc}^{\rm orb}$ to be $\sim$ 0.95 T/$\mu_{\rm B}$. 
In addition, the peak positions of the splitting NMR spectra at 4.2 K (Fig. \ref{peak_fit_NMR}) yield
\begin{eqnarray}
H_{{\rm int},c}^{\rm Si_{\rm I}} = 0.62 {\ }{\rm T}, {\ }{\ }H_{{\rm int},c}^{\rm Si_{\rm II}} = H_{{\rm int},c}^{\rm Si_{\rm III}} = 0.22 {\ }{\rm T}.
\end{eqnarray}
Thus, the $c$-components of Eqs. (\ref{eq:Hsite2}) are transformed into the following equations:
\begin{align}
\begin{aligned}
&m_{\rm ord}[4A_{cc}^{\rm U_1}-A_{cc}^{\rm U_2}] + 0.098 \ {\mu}_{\rm B} \times 0.95 \ {\rm T}/{\mu}_{\rm B} \\ 
& \ + m_{\rm ord} \times (-0.10 \ {\rm T}) = 0.62 \  {\rm T}, \\
&m_{\rm ord}A_{cc}^{\rm U_2} + 0.098 \ {\mu}_{\rm B} \times 0.95 \ {\rm T}/{\mu}_{\rm B} \\
& \ + m_{\rm ord} \times (-0.030 \ {\rm T}) = 0.22 \ {\rm T}.
\end{aligned}
\label{eq:Hsite1c}
\end{align}
In order to solve these equations, we need one more relation and will use the condition obtained in the paramagnetic state:
\begin{equation}
m_{c}^{\rm para}(4A_{cc}^{\rm U_1} + A_{cc}^{\rm U_2})
-0.041m_{c}^{\rm para}
=
m_{c}^{\rm para}A_{\parallel},
\label{eq:Hpara}
\end{equation}
where $m_{c}^{\rm para}$ is the induced magnetization for $H \parallel [001]$ and $A_{\parallel}$ ($\sim 0.69$ T/$\mu_{\rm B}$) is obtained from the $K$-$\chi$ plot, as already mentioned above.
The second term is the dipolar-field correction calculated in the manner described above.
From the Eqs. (\ref{eq:Hsite1c}) and (\ref{eq:Hpara}), we have finally obtained
$m_{\rm ord} \sim 1.4 \ {\mu}_{\rm B}$, $\ A_{cc}^{\rm U_1} \sim 0.15 \ {\rm T}/\mu_{\rm B}$, and $\ A_{cc}^{\rm U_2} \sim 0.12 \ {\rm T}/\mu_{\rm B}$. 
The estimated value of $m_{\rm ord}$ yields the saturation magnetization of about 0.47  $\mu_{\rm B}/{\rm U}$ ($ = 1.4 \times \frac{1}{3}$) in the assumed up-up-down structure.
On the other hand, from the magnetization measurements on single crystalline UAu$_2$Si$_2$, it was estimated to be approximately 0.25 $\mu_{\rm B}/{\rm U}$.
This discrepancy can be attributed mainly to imperfections of our simple model used in the analysis for the NMR spectra, such as the limited number of considered neighbors, neglects of the effects of diamagnetic fields and misorientation.

The present symmetry analysis assuming the $\bm{q}$ = (2/3, 0, 0) order gives a reasonable explanation also for the large broadening of the spectra observed below \tn \ in the in-plane external field. When $\bm{H}_{\rm ext} \parallel [100]$, the NMR spectra indeed depend on the [100] component of $\bm{H}_{\rm int}$. 
It can be easily noticed from the Eqs. (\ref{eq:Hsite2}) that we will observe three peaks in the NMR spectra if we use a single-crystalline sample. 
In the powdered sample, these three peaks will be observed as the large broadening of a peak due to the random orientation of the crystal in the basal plane, as observed in the present experiment.
Measurements on a single-crystalline sample are necessary to make a quantitative discussion about the non-diagonal components of the hyperfine coupling tensor.     

\section{Discussion}
It is notable that the same magnetic structure has been found in the so-called phase II of U(Ru$_{0.96}$Rh$_{0.04}$)$_2$Si$_2$, which is known to appear in a high magnetic field of $\sim$ 35 T applied parallel to [001] direction \cite{Kuwahara13}. 
More recently, the non-doped system URu$_2$Si$_2$ has also been revealed to exhibit a spin-density-wave order with a similar propagation vector, $\bm{q}$ = (0.6, 0, 0), in the phase II \cite{Knafo16}.
Because phase II is neighboring to the hidden order phase of URu$_2$Si$_2$, it may provide useful information about the hidden order to study in detail the AFM phase of UAu$_2$Si$_2$.
In particular, it will be interesting to examine what type of interactions provokes the system to order in this magnetic structure that breaks 4-fold symmetry in the $c$-plane. 
No other compound in the U$T_2$Si$_2$ family orders in a magnetic structure which breaks the in-plane rotational symmetry in zero magnetic fields.

In the above discussion, we simply assumed that the ordered moments are aligned in the [001] directions, although the present NMR data obtained using a powdered sample lacks specific information to evaluate the canting components.
This assumption was validated from our recent neutron diffraction experiments on a single-crystalline sample; the obtained results are consistent with the magnetic structure concluded from the present NMR study \cite{Tabata17-2}.
The canted component of the ordered moments is estimated to be less than 0.03 $\mu_{\rm B}$/U. We should, however, bear in mind that the uniaxial magnetic anisotropy of this compound originates mainly from the anisotropy in magnetic correlations, not from the single-ion crystalline-electric-field effects, as suggested by the almost isotropic effective $g$-value \cite{Tabata16}. Investigation of magnetic correlations of this compound through inelastic neutron scattering is in progress.


\section{Conclusion}
We have obtained the microscopic evidence for staggered order of magnetic moments in the ordered phase below \tn $\sim$ 20 K of UAu$_2$Si$_2$ through the $^{29}$Si-NMR experiments. From a peak-splitting pattern observed in the NMR signals, we have concluded that the spin-uncompensated AFM order with $\bm{q}$ = (2/3, 0, 0) would occur in the ordered phase.
The magnitude of the moment has been estimated to be $1.4 \pm 0.2 \ \mu_{\rm B}$ per uranium atom from the symmetry analysis.
The site-depending hyperfine-coupling constants have also been evaluated.
We detected non-negligible orbital contributions to the hyperfine fields, which may account for the linear component of the magnetization curve.
We have also pointed out that the magnetic structure proposed for UAu$_2$Si$_2$ is very similar to that observed in a high-field phase (Phase II) of U(Rh$_{1-x}$Ru$_x$)$_2$Si$_2$ for $x < 0.04$ implying the similarity in magnetic correlations underlying these compounds.

\begin{acknowledgment}
The present research was supported by JSPS Grants-in-Aid for Scientific Research (KAKENHI) Grant No. JP15H05882 and JP15H05885 (J-Physics), and for the Strategic Young Researcher Overseas Visits Program for Accelerating Brain Circulation (No. R2501). 
\end{acknowledgment}

\bibliographystyle{jpsj_tabata}
\bibliography{bib_UAS.bib}

\end{document}